# Evoking realistic affective touch experiences in virtual reality

Sofia Seinfeld[1†*], Ivette Schmidt[1], & Jörg Müller[1]

[1]Institute of Computer Science, University of Bayreuth, Bayreuth, Germany

† **Present address**: Centre de la Imatge i Tecnologia Multimèdia, Universitat Politècnica de Catalunya, Barcelona, Spain

\* **Corresponding author:**
Dr. Sofia Seinfeld
sofia.seinfeld@citm.upc.edu

## Abstract

This study aims to better understand the emotional and physiological correlates of being caressed in VR depending on the type of multisensory feedback provided and the animate or inanimate nature of the virtual representation that touches an embodied virtual body. We evaluated how pleasure, arousal, embodiment, and the illusion of being touched in VR were influenced by the inclusion of only visual feedback compared to visuotactile stimulation conditions, where participants, in addition to seeing an avatar or feather caressing their virtual bodies, also perceived congruent mid-air ultrasonic tactile stimulation or real interpersonal touch. We found that visuotactile feedback, either based on ultrasound or real interpersonal touch, boosts the illusion of being affectively touched and embodied in a virtual body compared to conditions only based on visual feedback. However, real interpersonal touch led to the strongest behavioral and emotional responses compared to the other conditions. Moreover, arousal and the desire to withdraw the caressed hand was highest when being touched by a female avatar compared to a virtual feather. Female participants reported a stronger illusion of being caressed in VR compared to males. Overall, this study advances knowledge of the emotional and physiological impact of affective touch in VR.

**Keywords:** Affective Touch, Virtual Reality, Embodiment, Haptics, Emotional Responses

## 1. Introduction

Touch is the first sense to develop in the human embryo and it is a vital communication channel in parent-infant interactions (Parsons et al., 2017). During all stages of life, tactile perception plays a prominent role in the exploration of the environment and in social bonding. For instance, touch enables us to perceive textures, pleasure, and pain, as well as to express emotions such as happiness, intimacy, love, insecurity, and threat (Andersen & Guerrero, 2008; Schirmer & Adolphs, 2017). Several studies have demonstrated the detrimental consequences of touch deprivation on physical and psychological health (Field, 2010; Jakubiak & Feeney, 2017). Despite the importance of touch in our everyday life, a significant amount of the worldwide population lacks meaningful and rich social interactions including interpersonal touch (Beßler et al., 2019). In fact, social isolation has become more evident with the COVID-19 crisis due to the required social distancing rules (Armitage & Nellums, 2020; Bavel et al., 2020). A potential solution to provide social touch experiences when real touch is not possible due to physical distance restrictions is through mediated social communication. The combination of immersive Virtual Reality (VR) systems with carefully designed tactile feedback is an ideal candidate for accomplishing interactions including social touch despite the artificial nature of the sensory stimuli used to achieve these illusions. Furthermore, recent studies have proposed that VR is an ideal medium to simulate and better understand interpersonal touch, since this type of touch frequently involves sensitive and intimate aspects that are impossible to replicate under laboratory conditions (Fusaro et al., 2021). However, until now there has been scarce research on how to achieve realistic affective touch in VR and its corresponding subjective and physiological correlates depending on the type of multisensory feedback delivered and the animate or inanimate nature of the virtual objects that touch an embodied artificial body (Haans & IJsselsteijn, 2006; Huisman, 2017; van Erp & Toet, 2015).

VR provides experimental settings with high ecological validity and internal control, making it an especially suitable technology to understand the role played by touch in different psychological, neurological, and social phenomena. For instance, de Jong et al. (2017) and Carey et al. (2021) investigated whether affective touch enhanced the full-body ownership illusion using VR or a physical mannequin, finding inconclusive evidence for both types of artificial bodies. These results stand in contrast with other studies showing that affective touch modulates the strength of the rubber hand illusion, where spatiotemporal congruent visuotactile stimulation leads to the perceptual illusion that a fake hand is part of the real body (Botvinick & Cohen, 1998; Crucianelli et al., 2013). In these studies, affective touch is defined as slow velocity (i.e., 1 to 10 cm/s) pleasant touch that is specifically coded by C-fiber tactile unmyelinated afferents (CT afferents) located in the non-glabrous skin, while non-affective touch is characterized by faster velocities of touch (Olausson et al., 2002). However, the notion that CT afferents are only present in non-glabrous skin has been challenged by the evidence provided by recent studies showing that CT afferents can be also found in glabrous non-hairy skin (McGlone et al., 2012; Watkins et al., 2021). Moreover, other models of affective touch have highlighted that the perception of touch as pleasant is not only modulated by bottom-up factors, such as CT afferents and sensory processing, but also by top-down factors which focuses on the cognitive interpretation of contextual social and affective cues related to the touch experience (Watkins et al., 2021). For instance, a top-down factor which might have influenced the results of some past studies related to affective touch is that frequently tactile stimulation is delivered using a brush and not based on direct skin-to-skin contact. In this regard, Kress et al. (2011) has shown that interpersonal touch compared to being touched by inanimate objects (e.g., a brush) results in different brain somatosensory and insular cortex responses. Therefore, it is still unclear whether affective touch modulates body ownership for an artificial body, and also how embodiment and arousal are influenced by the method and object used to deliver the touch (i.e., touch by a human or object).

In relation to arousal and pleasure feelings, Fusaro et al. (2019) capitalized on VR to investigate how painful and pleasurable tactile experiences were impacted by the visual perspective and cognitive stance of the perceiver. This study found that top-down factors modulate pleasure and pain perceptions, with first-person perspective (1PP) and a self-oriented cognitive stance leading to more intense touch experiences compared to adopting a third-person perspective and an others-oriented mental mode. In a further study, using a similar paradigm, Fusaro et al. (2021) found that mere observation of visually presented touch on an embodied virtual body is sufficient to evoke increased skin conductance responses (SCRs). Interestingly, the strength of these SCRs were dependent on factors such as the part of the virtual body being touched (e.g., touch on arms compared to touch on taboo body parts), participants' sexual orientation, and the gender of the touching avatar. Interestingly, in that study, high-level factors that modulate behavioral and physiological responses to social touch in real life, such as sexual orientation, also influenced responses to virtual touch in VR (Harjunen et al., 2017). Both of these studies exploited proprioceptive and visual congruency in relation to the real body in order to induce the illusion of interpersonal touch, i.e., seeing a collocated virtual body from 1PP being caressed or harmed by a needle. However, none of these studies included concomitant tactile feedback on the participants' real body, so it is also still unclear how affective touch in VR is influenced by the further inclusion of additional tactile feedback when compared to the mere observation of touch.

The main goal of the present study is to further investigate the principles underlying affective interpersonal touch in VR. We address whether there is a difference on perceived pleasantness, autonomic responses, and body ownership, when participants experience a life-size virtual body from a 1PP and are caressed either by another virtual agent or an inanimate object (i.e., virtual feather). Moreover, we also evaluate the strength of subjective and physiological responses to affective touch in VR when only the visual modality is exploited to simulate a virtual caress compared to the simultaneous presentation of visual and tactile feedback. Finally, we also assess whether ultrasonic mid-air haptic stimulation might be used to induce realistic affective touch illusions in VR in comparison to control conditions based on the real touch of a human or a feather. The decision of using ultrasound-based haptics to simulate affective touch was based on evidence showing that this type of feedback conveys various emotions (Obrist et al., 2015) and can be used to represent different tactile patterns on the hand (Paneva et al., 2020; Wilson et al., 2014).

## 2. Methods
### 2.1. Participants

A total of 24 participants took part in the study after providing signed consent (12 males; all right-handed, Mean Age=24.75, SD=3.19). All participants were German, except four (i.e., 3 Indian and 1 Ukrainian). Participants were students recruited from around the university campus. The exclusion criteria for this study was the intake of psychoactive medications, suffering from any type of neurological or sensory impairment, and having not corrected vision difficulties. This study was approved by the Ethics Committee of the XXXX and followed ethical standards according to the Helsinki Declaration. All participants received economic compensation for their participation. The experiment was carried out before the onset of the COVID-19 pandemic.

### 2.2. Experimental Design

This study was based on a 2x3 within-group experimental design. The first factor of the design was the *Type of Virtual Representation* with two levels: a) observing a virtual body experienced from 1PP being caressed on the hand either by a female avatar or b) by a virtual feather (**Figure 1 A–C**; see **Supplementary Video**). The second factor was the *Type of Multisensory Feedback* with three levels: a) the only visual touch condition, where participants only visually perceived a female avatar or virtual feather softly caressing the palm of their left virtual hand (**Figure 1A**), b) the ultrasound touch condition, where participants saw a female avatar or virtual feather touching the palm of their virtual hand, while simultaneously perceiving spatiotemporal congruent ultrasound-based tactile feedback on the palm of their real hand (**Figure 1B**), and c) the real touch control condition, where participants also saw a female avatar or feather touching the palm of their virtual hand, while simultaneously perceiving spatiotemporal congruent tactile feedback on the palm of their real hand delivered by a real human (i.e., experimenter touching the hand of the participants) or with a physical feather (**Figure 1C**).

The order of conditions was fully randomized for the factor *Type of Virtual Representation* (i.e., being caressed by an avatar or feather). With respect to the *Type of Multisensory Feedback* factor, we randomized the order of presentation of the visual touch condition (**Figure 1A**) and the ultrasonic mid-air haptic stimulation condition (**Figure 1B**), however the real touch condition was always presented at the end of the experiment. Since the real touch condition consisted in the perception of real skin-to-skin contact or touch with a physical feather (**Figure 1C**), we considered that it was important to always present it at the end of the experiment based on two main reasons: 1) to serve as a baseline of the subjective feelings and physiological reactions evoked by the most realistic experience the participants could experience, while still taking into account the virtual and immersive nature of the scene; 2) to avoid potential order effects based on the influence and judgment bias that the real touch condition could have on the rest of conditions which tried to simulate real touch in an artificial way (Schwarz, 1994). However, in the real touch condition, we also randomized the order of presentation between the avatar and feather conditions. Each experimental condition was separated by a short break, where participants took off the head-mounted display (HMD) and were requested to answer questions related to the experimental condition that they had just experienced.

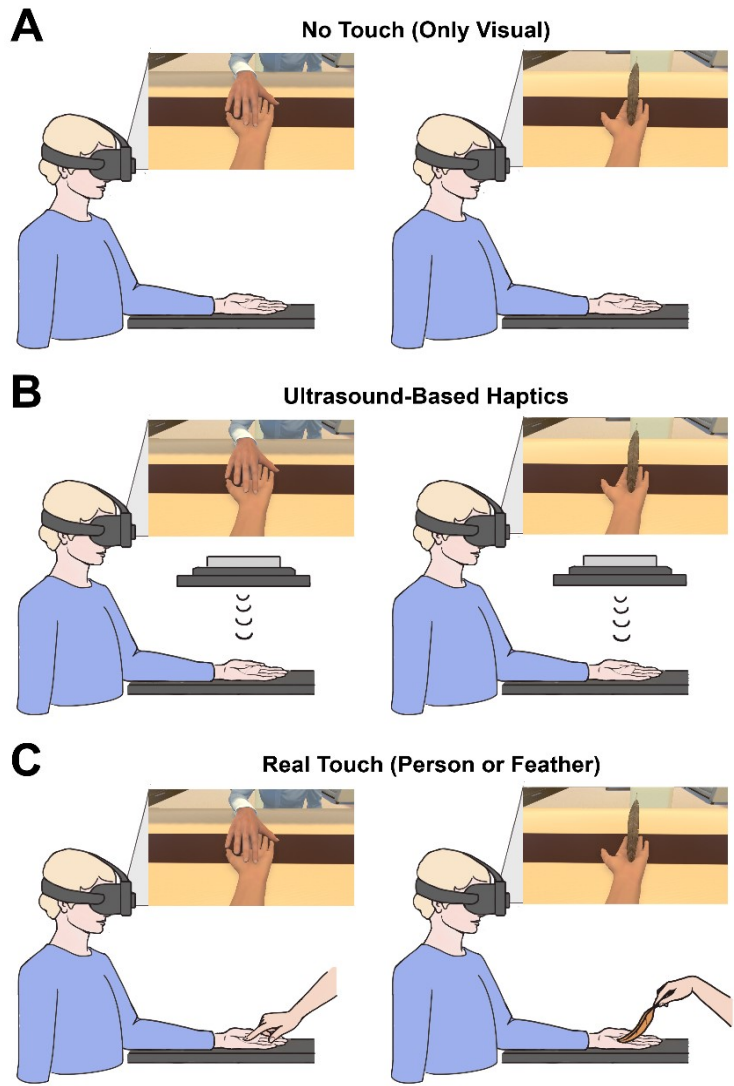

*Figure 1.* A) In the only visual touch condition, participants saw that their virtual body was being touched by either a female avatar or a virtual feather, while no concomitant tactile feedback was delivered; B) In the ultrasonic mid-air haptic stimulation condition, participants saw that their virtual body was being touched by a female avatar or a virtual feather, while simultaneously perceiving congruent ultrasound-based tactile feedback on their real hand; C) In the real touch condition, participants saw that their virtual body was being touched by a female avatar or feather while the experimenter caressed the participant's hand at the same time and location either with her hand or a physical feather.

### 2.3. Measures

#### Bodily and Touch Illusion Questions

We included questions to evaluate different aspects related to bodily and touch illusions in the immersive virtual environment, which are summarized in Table 1. The questions were answered on a 7-point Likert scale, where 1 meant "completely disagree" and 7 "completely agree" with the statement. The questionnaire items *Ownership*, *Control*, *Agency*, and *SelfLoc* focused on measuring the degree that participants felt embodied in the virtual body, hence experiencing the illusion that the virtual body was their own real body (Gonzalez-Franco & Peck, 2018; Serino et al., 2013). Furthermore, the questionnaire items *Caressed*, *Touched*, *Withdraw*, *TouchLocation*, and *TouchLocation (Control Question)* aimed to assess to what degree participants experienced the illusion of being caressed and touched in VR by the avatar and feather in the different experimental conditions. Similar subjective reports were used by Fusaro et al. (2016) and (2021). The detailed questionnaire items are listed in Table 1.

*Table 1. Questionnaire items asked to participants after each experimental condition*

| **Bodily Illusions Questions** | |
|---|---|
| *Ownership* | I felt as if the virtual hand I saw when I looked down was my hand. |
| *Control* | It felt like I could control the virtual hand as if it was my own hand. |
| *Agency* | The movements of the virtual hand were caused by my movements. |
| *SelfLoc* | I felt as if my hand was located where I saw the virtual hand. |
| **Touch Illusion Questions** | |
| *Caressed* | I felt as if my hand had been caressed. |
| *Touched* | It seemed as if the touch I felt was caused by the avatar/feather touching my virtual hand. |
| *Withdraw* | When the avatar/feather stroked me, I felt the instinct to withdraw my hand. |
| *TouchLocation* | It seemed as if I felt the touch of the stroking hand/feather in the location where I saw my virtual hand being touched. |
| *TouchLocation (Control Question)* | It seemed as if the touch I felt was located somewhere between my physical hand and my virtual hand. |

### Pleasantness and Arousal Rating Scale

After each experimental condition, participants also rated the *Pleasantness* of the touch using a scale from 0 (not at all pleasant) to 100 (extremely pleasant). They also rated the degree to which they judged the touch as *Unpleasant* from 0 (not at all unpleasant) to 100 (extremely unpleasant) and as *Arousing* with a scale from 0 (not arousing at all) to 100 (extremely arousing). Similar rating scales were used by Crucianelli et al. (2013).

### Skin Conductance

Skin conductance is a widely used measure of emotional arousal, which directly depends on sweat glands innervated by the autonomic nervous system (Critchley, 2002; van Dooren et al., 2012). SCRs have been used in past studies to assess threat-related responses towards an embodied mannequin or virtual body (Petkova & Ehrsson, 2008; Tieri et al., 2015). However, SCR has also been used to measure arousal in response to being visually touched by an avatar by Fusaro et al. (2021). In their study, they found that seeing a virtual body being touched in intimate areas (i.e., genitals or breasts) from 1PP leads to higher SCR when compared to being touched in neutral or less erogenous virtual body areas. The authors attribute this enhanced physiological reactivity to the erogenous value given to the virtual touch on intimate body parts. In the present study, we further evaluated how SCRs are modulated depending on being touched either by an avatar or inanimate object (feather) and depending on the type of multisensory feedback provided during the virtual touch experience. For instance, there is evidence that human touch leads to enhanced SCR compared to being touched by an object or not being touched at all (Hazem et al., 2018). In the present study, we aim to further assess whether these results are also replicated in VR. Moreover, Etzi & Gallace, (2016) found that tactile presentation of textures leads to higher SCR compared to the mere visual representation of textures. Thereafter, through this study we also aimed to analyze whether arousal differed depending on the inclusion of only visual feedback compared to conditions including visual and tactile feedback either based on ultrasound or real touch.

### 2.4. Virtual Reality Scene, avatar animation, and ultrasound haptics calibration

The virtual scene was programmed using Unity 3D and experienced through an HTC Vive HMD. The immersive virtual environment consisted of a room with chairs, table, television, sofa, window, and a couple of paintings. During the experiment, participants were requested to remain seated in front of a physical table, while simultaneously seeing a virtual body from 1PP sitting down in front of a virtual table in the same position. All participants were embodied in the same avatar, which was neutrally dressed (i.e., wearing a pair of jeans and t-shirt) in order to minimize the presence of prototypical features related to either the male or female gender. We also enabled hand and finger tracking through a Leap Motion sensor in order to enhance the sense of embodiment through visuomotor correlations at specific parts of the experiment, i.e. the virtual hands moved according to the movements executed by the participants with their real hands.

We animated the caresses of the female avatar or feather based on real-time motion capture using a PhaseSpace Impulse X2E, MotionBuilder, and Blender. The animation consisted of either a female avatar or feather softly caressing the palm of the left hand of the avatar that participants embodied from a 1PP during the experiment. More specifically, at the start of each experimental condition participants saw their virtual left hand lying on a virtual table, while their real physical left hand was lying in the same position on a physical table. In the same virtual room, in the female avatar conditions, they saw a virtual woman sitting on the opposite side of the table. The face of the female avatar was covered with a virtual translucent glass. Since the translucent glass was placed at the height of the chest of the female avatar, it was still possible to directly see the female avatar's hands and body, but not her face. Therefore, participants could see the female avatar extending her right hand to caress them, however it was impossible to see her facial expressions. We opted to use this translucent glass to hide the avatar's face, because we were primarily interested in the impact of social touch (i.e., caressing), without the additional influence of emotional cues provided by facial expressions which play a role in how touch experiences are judged in VR (Harjunen et al., 2017). Furthermore, we also wanted to avoid the possible occurrence of an uncanny valley effect related to the feelings of revulsion or uneasiness that might arise when a humanoid avatar imperfectly resembles an actual human being (Shin et al., 2019). In the feather condition, instead of an avatar, participants saw a virtual feather floating in mid-air and slowly moving while touching the palm of their left hand. See **Supplementary Video** for the VR scenarios.

We decided to deliver the touch on the palm of a participant's left hand in the different experimental conditions, since this body area is more sensitive to the perception of ultrasound waves due to the presence of high-density A-beta fibers which are critical for discriminative touch, but also because recently it has been shown that CT afferents, which are important for affective and affiliative touch, are also present in this body area (Watkins et al., 2021). For the experimental conditions where tactile feedback was based on ultrasound, we previously calibrated the ultrasonic waves according to the size of the participant's hand. This was done in order to avoid mismatches between the seen virtual touch and the felt ultrasonic tactile feedback on the palm of the hand. For this, we created a reference image of the left hand, where three positions were marked to illustrate the trajectory that the haptic feedback should follow, based on the position where the animation of the avatar and the feather started and ended caressing the palm of the virtual hand. This calibration phase was performed before participants entered the VR scene, without revealing any details of the actual experiment, therefore at this point participants were ignorant of the purpose of perceiving ultrasound on their hands. At this stage, we only requested that participants place their left hand on top of a table with the palm of their left hand facing upwards. An ultrasound board was built on top of the table in order to stand above the participants' palms. The structure to hold the ultrasound board on top of the table was specifically designed and built for this experiment using a 3D printer (see **Figure 2**). Once participants had their left hand lying on the table in the correct position, we started with the calibration process by making participants perceive two points of tactile stimulation on their left palm. These two points corresponded to the start and end points of the stimulation. Importantly, at this calibration stage, participants were able to control and move the location where they felt this tactile stimulation through the use of keys on a keyboard that they controlled with their right hand. We requested that participants match the location where they felt the two points of tactile stimulation on the palm of their real hand, with the start and end reference points that they observed in the reference image of a left hand. Once participants reported that the location of the perceived tactile feedback matched with the reference image, we recorded the locations of this reference points in order to define the trajectory that the ultrasound waves should follow in order to be synchronized with the visual animations of the avatar and feather. These positions were transformed into velocities by subtracting their value from the lower point that was felt during the emission of the two simultaneous points. These velocities varied depending on the palm size since the stroking animation was always 2.3 seconds long, so for instance assuming that the palm size of a specific participant was 10 cm, the velocity of the stroking was approximately 4.3 cm/s, and for a palm size of 9 cm, the velocity of the stroking was 3.9 cm/s. Overall, stroking velocities always ranged between 1–10cm/s, considered to be optimal velocities to stimulate C-fiber tactile (CT) afferents which are important receptors in affective touch experiences (Olausson et al., 2002).

The results of this calibration were transmitted to the ultrasound board through the Unity console, and the calibration reference points could be copied and pasted to adjust the emission velocities accordingly. Ultrasound intensity, frequency, and distance were parameters that we kept constant throughout the different experimental conditions. Despite the start and end location values of the ultrasonic haptics being adapted to participant hand size, these values did not vary much between individuals, and they provide a good synchronization of haptics and visuals based on a pilot study we performed before running the actual experiment.

SCRs was measured using a ProComp Infiniti Encoder (Thought Technologies Ltd) connected to a PC through a serial port and based on the placement of two galvanic skin response sensors (i.e., SA9309M bipolar electrodes) on the middle and index fingers of a participant's right hand. Participants used the SCR electrodes throughout all the experiments. The signal was sampled at 256 Hz and low-pass filtered with a cut-off frequency of 1 Hz. The data was preprocessed using MATLAB and analyzed using SPSS. Before the analysis, SCR data was visually inspected to detect possible undesired noise and artifacts due to motion. The experiment was programmed to send an event marker to the SC signal as soon as either the avatar or feather started caressing the participant for 60 seconds in each experimental condition.

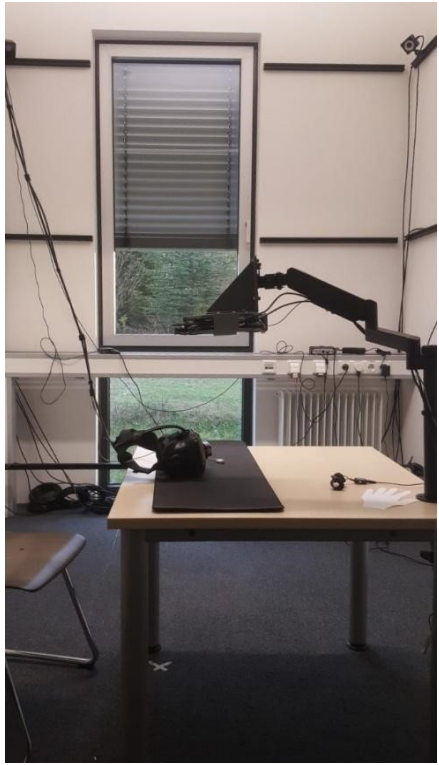

*Figure 2. Image of the equipment used for the experiment. In addition to the head-mounted display which had a Leap Motion attached in front of precise hand tracking, we also 3D printed a special structure to mount on top of the table in order to place the ultrasound board that delivered the haptic feedback in some experimental conditions.*

### 2.5. Procedure

Upon arrival in the laboratory, participants were given basic information about the study and signed a consent form if they were willing to participate. Participants were requested to sit at a table during the study. For all the experiments, we decided to have the same female experimenter executing the study, in order to match the gender of the avatar that caressed the participants. Before starting with the actual experiment, we followed the procedure described in the previous section to calibrate the ultrasound-based haptics according to the size of the participant's left palm. SCR sensors were then placed on the middle and index fingers of the participant's right hand. Finally, participants were requested to wear a HMD with a Leap Motion sensor attached to it.

When the VR scene started, participants saw themselves immersed in a virtual living room that had a virtual table located in front of them. They also saw a virtual counterpart of their hands, which moved according to their real hand movements. To familiarize participants with the virtual scene, we first asked them to briefly describe their surroundings (i.e., the virtual room) and their new virtual bodies. Here we explicitly asked participants to look down at their virtual body and to move their virtual hands and look at them. Subsequently, we asked participants to relax, remain silent, and breathe slowly for 2 minutes, to record a baseline measure of their physiological state. When the recording of the physiological baseline ended, the scene vanished (i.e., turned black).

Before the start of the first experimental condition, with the HMD screen still blacked out, the experimenter took the participant's real left hand, and placed it in the same position used in the calibration phase of the ultrasonic haptics, based on a drawing of the silhouette of the real hand performed in that phase on top of the physical table. Therefore, in all the experimental conditions, the left hand was always placed in the same position. We asked participants to closely look at their left hand when the scene appeared again. When the scene reappeared, depending on the experimental condition, participants either saw a virtual feather or female avatar caressing their hand. The stimulation lasted a total of 60 seconds, and then the scene vanished again (i.e., turned black). When each experimental condition ended, there was a short break where participants took the HMD off and filled out questionnaires on a tablet device about the VR experimental condition they just had experienced. This same procedure was followed for all the experimental conditions. Before each experimental condition started, participants were requested to closely look at their left hand, something that the experimenter could verify through a screen where she could see the participant view of the VR scene. So in all experimental conditions, we ensured that participants were paying attention to the different types of stimulation.

### 2.6. Statistical Analysis

Embodiment and affective valence questionnaire items were analyzed using multilevel mixed-effects ordered logistic regressions using Stata 16.1 software. In the multilevel ordered logistic regressions, *Type of Multisensory Feedback* (i.e., only visual touch, ultrasound touch, and real touch condition), *Type of Virtual Representation* (i.e., avatar or feather), and *Gender*, as well as their interaction terms, were introduced as fixed factors. Participants' IDs were set as random effects in the model in order to control for the within-group nature of the experiment. We decided to include the additional factor *Gender* in the analysis to control for the potential influence of participants being either male or female in the results.

SCR data was first tested for normality using a Shapiro-Wilk test. Based on the normal distribution of the data, a 2x3 repeated measures ANOVA with factors *Type of Virtual Representation* (i.e., avatar or feather) and *Type of Multisensory Feedback* (i.e., only visual touch, ultrasound touch, and real touch condition) was performed. Moreover, we also included *Gender* (i.e., female and male) as a between-group factor in the analysis. In all the analysis carried out, *Gender* was subsequently excluded from the analyses if it yielded no significant differences.

### 3. Results

#### Embodiment Questions

The results indicate that the inclusion of visuotactile feedback enhances the feeling of embodiment when compared to conditions that only included mere visual stimulation. Specifically, we found significantly higher body ownership (i.e., *Ownership*) scores when tactile feedback was delivered either through real touch ($z=5.66$, $p<0.01$, CI=1.59 – 3.27) or ultrasound haptics ($z=3.02$, $p<0.01$, CI=0.40 – 1.88) compared to conditions only based on the perception of visual feedback. The same was true for self-location (*SelfLoc*), where participants felt a higher sense of being located within the virtual body in the conditions including real tactile feedback ($z=3.48$, $p<0.01$, CI=0.77 – 2.77) and ultrasonic mid-air haptic stimulation ($z=2.87$, $p<0.01$, CI=0.44 – 2.36), compared to the only visual touch condition. Body ownership was also significantly higher in the real touch condition than in the ultrasonic haptics condition ($z=3.18$, $p<0.01$, CI=0.49 – 2.08). The sense of *Control* of the virtual hand was significantly higher in the real touch condition compared to the only visual touch condition ($z=3.19$, $p<0.01$, CI=0.53 – 2.21). Similarly, the perception of *Agency* was significantly higher in the ultrasound haptics condition compared to the only visual touch condition ($z=2.37$, $p=0.02$, CI=0.20 – 2.07).

No significant main effect of *Type of Virtual Representation* (i.e., avatar or feather) was found in *Ownership*, *Control*, *Agency* and *SelfLoc*. Moreover, no significant effect for the *Gender* was found, indicating that the gender of the participants does not seem to play a role in embodiment related questions. Detailed results can be seen in **Table 2** and **Figure 3**.

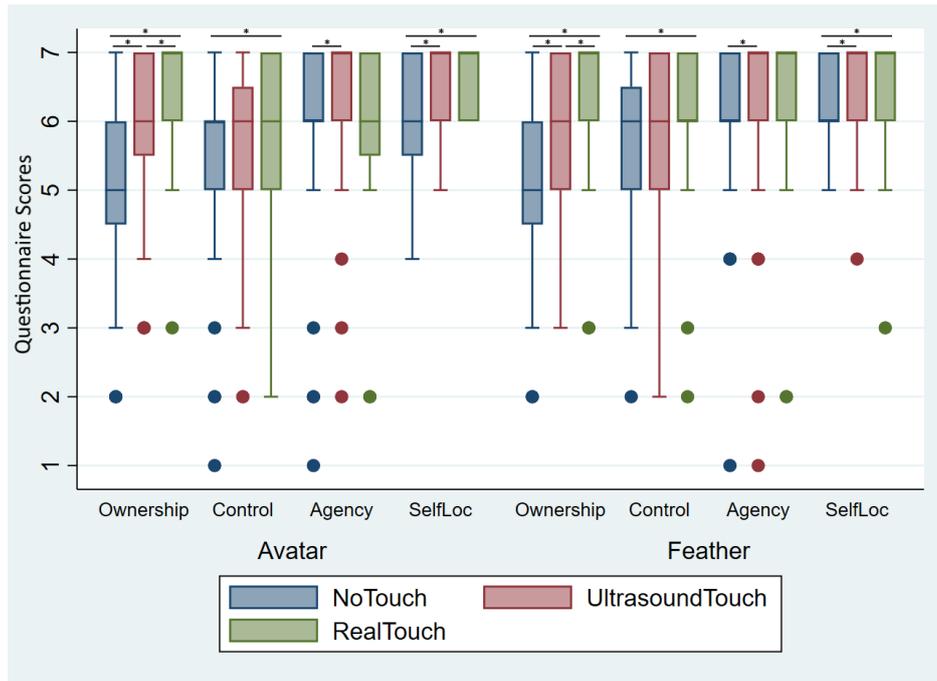

Figure 3. Boxplots of the different embodiment related questions.

Table 2. Results of the multilevel order logistic regression on the embodiment questions related to the type of multisensory feedback delivered.

|  | RealTouch vs NoTouch | RealTouch vs UltrasoundTouch | UltrasoundTouch vs NoTouch |
|---|---|---|---|
| *Ownership* | z=5.66, p<0.01, CI=1.59-3.27 | z=3.18, p<0.01, CI=0.49-2.08 | z=3.02, p<0.01, CI=0.40-1.88 |
| *Control* | z=3.19, p<0.01, CI=0.53-2.21 | z=1.51, p=0.13, CI=-0.17-1.30 | z=1.13, p=0.258, CI=-.34-1.28 |
| *Agency* | z=1.55, p=0.12, CI=-.19-1.59 | z=-0.76, p=0.45, CI=-1.08-0.47 | z=2.37, p=0.02, CI=0.20-2.07 |
| *SelfLoc* | z=3.48, p<0.01, CI=-0.77-2.77 | z=0.66, p=0.51, CI=-0.55-1.11 | z=2.87, p<0.01, CI=-0.44-2.36 |

## Touch Illusion Questions

In relation to the perceptual illusion of being touched by an avatar or feather, we have found that the inclusion of additional tactile feedback clearly enhances the illusion of being caressed and touched compared to the inclusion of mere visual feedback. Specifically, participants reported a stronger illusion of being *Caressed* by the avatar and feather when real (z=8.79, p<0.01, CI=5.31 – 8.35) or ultrasonic mid-air haptic stimulation (z=6.66, p<0.01, CI=2.35 – 4.30) were included compared to the only visual condition. The same was true for the *Touched* question where higher scores were reported for the real touch (z=6.18, p<0.01, CI=1.93 – 3.72) or ultrasound-based haptics (z=4.19, p<0.01, CI=0.87 – 2.39) conditions, compared to the only visual condition. However, still a significantly higher illusion of being *Caressed* (z=5.68, p<0.01, CI=1.94 – 3.98) and *Touched* (z=2.70, p<0.01, CI=0.29 – 1.85) was reported in the real touch condition compared to the ultrasound-based tactile condition. Participants also reported a significantly higher desire to withdraw their real hand when the avatar or feather was touching the virtual hand in conditions including real tactile feedback compared to the condition including only visual feedback (z=2.75, p<0.01, CI=0.35 – 2.11). Moreover, participants also indicated a stronger illusion of being touched in the location where they saw the touch on the virtual hand in those conditions including real touch (z=7.27, p<0.01, CI=2.71 – 4.72) or ultrasound tactile feedback (z=6.41, p<0.01, CI=2.07 – 3.90) compared to the only visual condition. Details of the results found in the *Type of Multisensory Feedback* can be seen in **Table 3**.

In addition to the aforementioned result, we also found a main effect of *Gender* only in the *Caressed* question and of *Type of Virtual Representation* (i.e., avatar or feather) in the *Withdraw* question. Female participants reported higher scores in the illusion of being *Caressed* in VR compared to male participants (z=2.44, p=0.01, CI=0.24-2.19), independently of the type of feedback delivered. Moreover, in the different experimental conditions participants also reported a stronger desire to *Withdraw* their hand when touched by an avatar compared to a virtual feather (z=-3.54, p<0.01, CI=-2.05-0.59). No other significant differences in relation with *Gender* and *Type of Virtual Representation* were found. See **Figure 4** for more details.

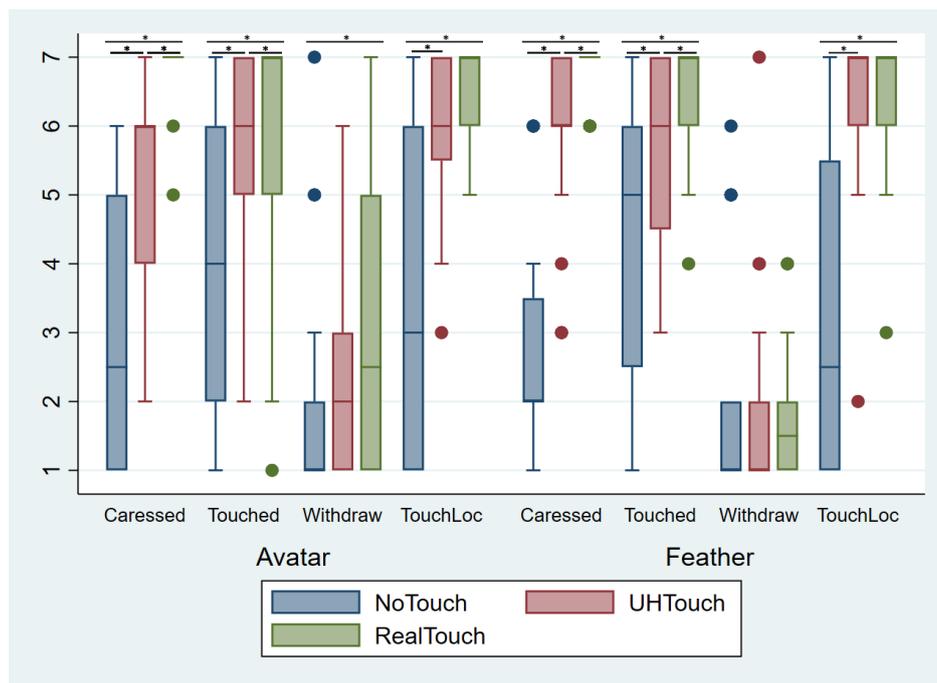

*Figure 4. Boxplots of the touch related questions.*

*Table 3. Results of the multilevel order logistic regression on the touch related questions in relation to the type of multisensory feedback delivered.*

|  | **RealTouch vs NoTouch** | **RealTouch vs UltrasoundTouch** | **UltrasoundTouch vs NoTouch** |
|---|---|---|---|
| *Caressed* | z=8.79, p<0.01, CI=5.31-8.35 | z=5.68, p>0.01, CI=1.94-3.98 | z=6.66, p<0.01, CI=2.35-4.30 |
| *Touched* | z=6.18, p<0.01, CI=1.93-3.72 | z=2.70, p<0.01, CI=0.29-1.85 | z=4.19, p<0.01, CI=0.87-2.39 |
| *Withdraw* | z=2.75, p<0.01, CI=0.35-2.11 | z=1.08, p=0.28, CI=-0.33-1.14 | z=1.48, p=0.14, CI=-0.22-1.55 |
| *TouchLocation* | z=7.27, p<0.01, CI=2.71-4.72 | z=1.69, p=0.09, CI=-.11-1.45 | z=6.41, p<0.01, CI=2.07-3.90 |
| *TouchLocation (Control Question)* | z=-1.64, p=.10, CI=-1.48-0.13 | z=-1.40, p=0.16, CI=-1.27-0.21 | z=0.06, p=0.95, CI=-0.75-0.80 |

## Affective Valence Questions

Overall, conditions that included visuotactile feedback compared to conditions that only included visual feedback resulted in increased *Pleasantness* and *Arousal*. We found significantly higher *Pleasantness* scores in the real (z=5.50, p<0.01, CI=1.45-3.06) and ultrasonic mid-air haptic stimulation (z=2.51, p=0.01, CI=0.20-1.66) conditions compared to the only visual condition. Moreover, reported *Pleasantness* was higher in the real touch condition compared to the ultrasonic mid-air haptic stimulation. Similar results were found in the *Arousing* rating scale, with significantly higher scores for the real (z=7.26, p<0.01, CI=0.96-2.55) and ultrasound haptics condition (z=4.32, p<0.01, CI=0.96-4.29) compared to the only visual condition. However, higher *Arousing* responses were also found for the real touch condition compared to the ultrasonic haptics condition (z=3.01, p<0.01, CI=0.38-1.81). No significant differences were found in the *Unpleasantness* rating scale. Details of the results found in the Type of Multisensory Feedback can be seen in **Table 4**.

We also found a main effect of *Type of Virtual Representation* in *Arousing* (z=-2.08, p=0.04, CI=-1.26- -0.36), showing that higher arousal scores were reported when being touched by a female avatar compared to a virtual feather. No main effect of *Gender* was found in any of these rating scales. See **Figure 5** for more details.

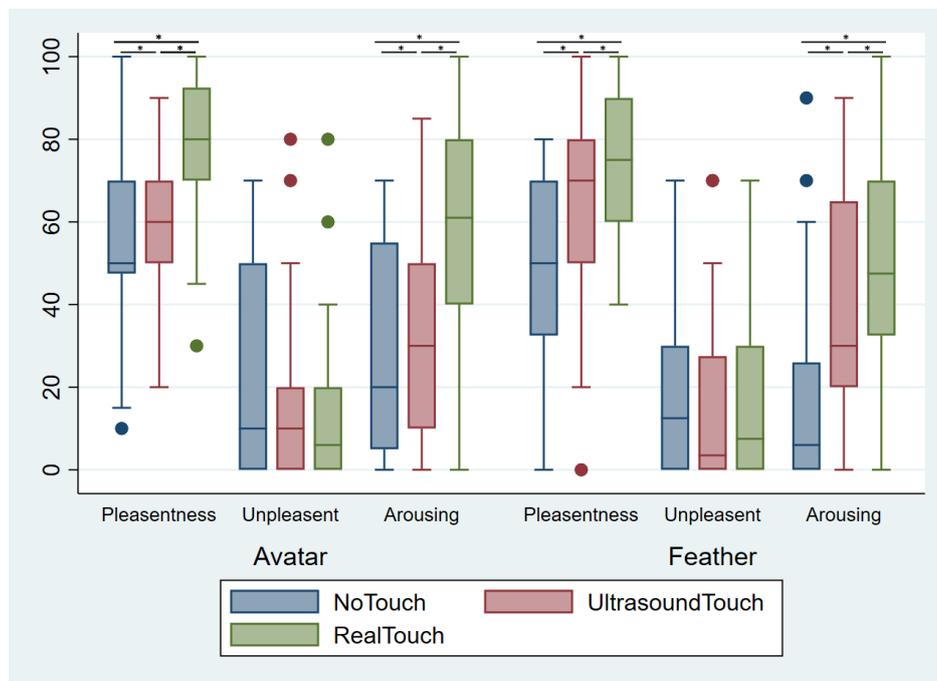

*Figure 5. Boxplots of the affective valence questions.*

Table 4. Results of the multilevel order logistic regression on the affective valence questions in relation to the type of multisensory feedback delivered.

|  | RealTouch vs NoTouch | RealTouch vs UltrasoundTouch | UltrasoundTouch vs NoTouch |
|---|---|---|---|
| *Pleasantness* | z=5.50, p<0.01, CI=1.45-3.06 | z=3.19, p<0.01, CI=0.46-1.92 | z=2.51, p=0.01, CI=0.20-1.66 |
| *Unpleasantness* | z=-1.23, p=0.22, CI=-1.29-0.30 | z=-0.19, p=0.85, CI=-0.78-0.65 | z=-0.47, p=0.64, CI=-0.96-0.59 |
| *Arousing* | z=7.26, p>0.01, CI=0.96-2.55 | z=3.01, p>0.01, CI=0.38-1.81 | z=4.32, p>0.01, CI=0.96-4.29 |

## Physiological Reactions

A repeated measures ANOVA indicated that there was a significant main effect of the *Type of Multisensory Feedback* ($F(2,22)=7.70$, $p<0.01$, $np2=0.41$), showing that those conditions that included tactile feedback (i.e., real touch or ultrasound haptics) resulted in a higher percentage of change in SCRs compared to the only visual touch condition (i.e., NoTouch) which did not include additional tactile feedback as can be seen in **Figure 6**. Moreover, we also found a significant interaction between *Type of Multisensory Feedback* and *Type of Virtual Representation* ($F(2,22)=6.27$, $p<0.01$, $np2=0.36$), indicating that the real touch condition led to a stronger physiological response denoted by higher changes in SCRs when participants were touched by a female avatar and simultaneously touched by the female experimenter compared to seeing the touch of a virtual feather and simultaneously feeling the touch of a physical feather (**Figure 6**). Interestingly, these significant differences were not observed when tactile feedback was based on ultrasonic mid-air haptics or in the condition only based on visual feedback (i.e., only visually seeing the touch).

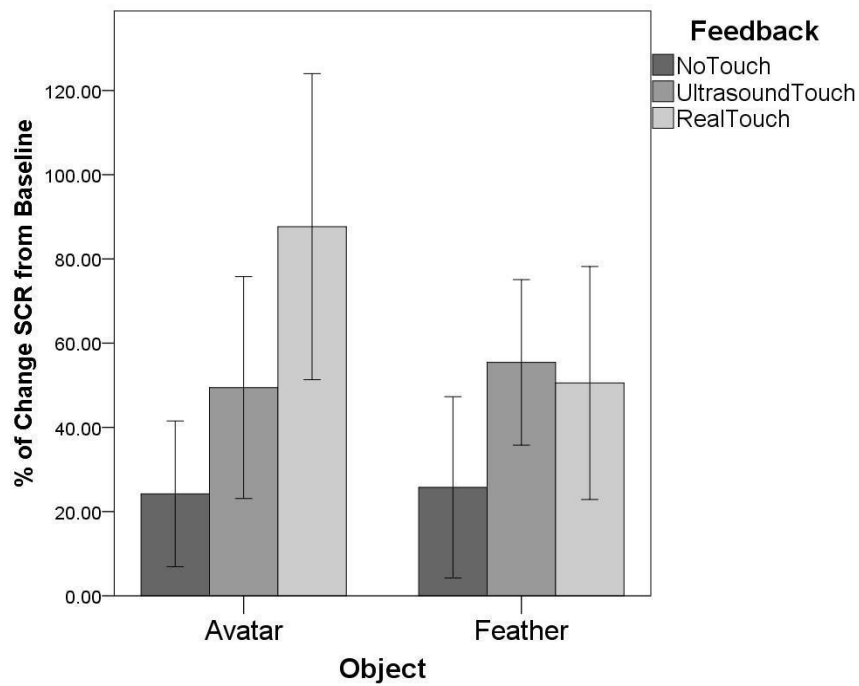

*Figure 6. Percentage of change of Skin Conductance Responses (SCR) during the first 30 seconds of each experimental condition.*

## 4. Discussion

The present study further contributes to a better understanding of some of the emotional and physiological correlates of simulating affective touch in immersive VR. Our results indicate that the inclusion of congruent spatio-temporal visuotactile feedback, compared to mere visual stimulation, boosts the illusion of being affectively touched and embodied in an artificial body in VR, as well as increases feelings of pleasantness, arousal, and physiological responses. Importantly, we show that bodily related illusions (i.e, perception of owning a virtual body or being caressed) are enhanced either by the presence of real interpersonal touch (i.e., skin-to-skin or physical feather contact) or by the simulation of touch using ultrasonic mid-air haptic stimulation. However, the condition based on real interpersonal skin-to-skin contact while seeing a female avatar touching the embodied virtual body always led to the maximal level of experienced illusory affective touch (i.e., being caressed), embodiment, and physiological response. In line with past evidence, these results highlight the critical role played by somatosensory stimuli when evoking bodily-related and affective touch illusions (Crucianelli et al., 2013; Maselli & Slater, 2013, 2014). The present study also underscores a series of challenges faced when aiming to replicate realistic social touch in simulated virtual immersive environments (van Erp & Toet, 2015). In the following paragraphs, we discuss in more detail the implications and results of the present study with respect to relevant past evidence.

One of the main findings of the present study is that the illusion of being caressed and the experienced level of arousal and pleasantness in response to being touched by an avatar or virtual feather were enhanced with the inclusion of tactile feedback. Past research assessed the impact of seeing a virtual body from 1PP being affectively touched by other avatars, however these studies did not include the delivery of simultaneous tactile stimulation on participants' real bodies, while concurrently observing touch on the embodied virtual body. Interestingly, these past studies found that the single observation of touch on an embodied body was enough to induce vicarious feelings of somatosensory stimuli on one's body, which evoked different emotional responses depending on high-order factors (Fusaro et al., 2019; Fusaro et al., 2021). For instance, Fusaro et al. (2021) showed that the erogenous value attributed to being visually touched in VR was modulated by participants' sexual orientation (e.g., being touched by a female or male avatar) and the area of the virtual body being touched (Fusaro et al., 2021). Moreover, the single observation of a painful (i.e., injection with a needle) or pleasurable stimulus (i.e., caress) exerted to a virtual body seen from 1PP led to strong emotional and physiological responses (Fusaro et al., 2019). The results of the present study partially support these past findings, since despite also observing that the single observation of virtual touches evoked changes in emotional response, these reactions were significantly more vivid and intense in the experimental conditions that included additional tactile stimulation, as evidenced by subjective reports and physiological responses. These results are in line with a past study that found significantly higher SCR during the tactile presentation of textures rather than their mere visual presentation (Etzi & Gallace, 2016). The enhanced emotional responses found in the present study when including visuotactile feedback compared to only visual feedback is an important aspect to take into account in future studies when trying to simulate social touch experiences in VR, and also in studies that exploit VR as a medium to better understand real-world social interactions that include interpersonal touch.

Another interesting finding is that despite participants reporting a strong illusion of being touched either by a female avatar (i.e, humanoid representation) or feather (i.e., inanimate object), the desire of participants to withdraw their virtual hand and their level of experienced arousal was higher when they were caressed by a female avatar in comparison to being touched by a virtual feather. This result was confirmed through more objective physiological measures, where the highest level of SCR was elicited by being touched by a female avatar in the visuotactile conditions, especially in the ones based on real skin-to-skin touch, compared to the other experimental conditions. Interestingly, in a past study that addressed the impact of the simultaneous presentation of visual and tactile stimuli, Etzi et al. (2018) did not find significant differences in electrodermal responses between skin-to-skin contact and textures during the concurrent observation of images with different visual emotional valences. Nevertheless, the authors argued that the lack of differences between conditions was potentially related to the artificial nature of the experimental setup, which was far from the natural conditions of social interaction. In the present study, where VR was used to simulate scenes including touch interactions with high ecological validity, we did find that skin-to-skin contact compared to other conditions, as well as the addition of visuotactile stimulation compared to only visual feedback, leads to an increased emotional and physiological response. On a neural level, this finding could potentially be related to larger somatosensory and insula responses evoked by skin-to-skin contact, compared to being touched with inanimate objects (Kress et al., 2011), an effect which also seems to occur in immersive virtual environments, based on the present findings.

In terms of how embodiment is modulated by being affectively touched in VR, we also found that the inclusion of visuotactile feedback compared to only visual feedback resulted in the highest levels of embodiment. Higher levels of embodiment were reached with respect to the only visual conditions, despite the tactile feedback being based on ultrasonic mid-air haptics or real interpersonal touch. However, higher levels of embodiment were reported in the real interpersonal touch conditions compared to the conditions where touch was simulated using ultrasonic airwaves. The degree of body ownership, self-location, and agency experienced did not differ significantly between being touched by a humanoid virtual representation (i.e., female avatar) or virtual feather, suggesting that the type of virtual representation that caressed the embodied virtual body did not play a differential role in the degree of experienced embodiment. In a past study, Fusaro et al., (2019) found similar results, since body ownership was enhanced when seeing a caress of a virtual body experienced from 1PP, an effect not observed with painful or neutral stimuli. However, no additional tactile feedback was included in this study and touch was simulated only exploiting the visual modality. The present research further shows that although embodiment is already enhanced with the mere observation of affective touch, this effect is more evident when additional somatosensory stimuli are included in the VR experience. Another interesting aspect is that Fusaro et al., (2019) stressed the importance of using natural human kinematics to emulate the caresses, since in a previous study in which they used unnatural preprogrammed avatar movements to simulate a caress, embodiment was not enhanced (Fusaro et al., 2016). This is an important aspect that deserves further research, since in the present study we also took special care in animating the movement of the avatar and feather using full body motion tracking recordings, and replicated the earlier results obtained by Fusaro et al. (2019) with respect to embodiment. Therefore, it is possible that the naturalness of the motion of an avatar that interacts with us in VR also plays a role in how we perceive our virtual body and environment.

Despite the findings summarized above showing the potential role of affective touch in enhancing the sense of body ownership, evidence so far in relation to this topic is inconclusive, possibly due to the use of different definitions of affective touch and experimental setups. For instance, while Crucianelli et al. (2013) findings suggested that affective touch enhanced illusory ownership of an artificial rubber hand compared to non-affective touch, de Jong et al. (2017) did not replicate this effect when evoking a full body ownership illusion in VR, nor did Carey et al. (2021) when evoking a body ownership illusion towards a mannequin body. Importantly, in these studies affective touch was defined as slow touch (1-10 cm/s) applied to non-glabrous hairy skin, while non-affective touch was defined as touch administered with faster velocities. Interestingly, these studies also found that sensory and spatial congruency seem to be more critical for strengthening body ownership illusions when compared to the affective components of touch (Crucianelli et al., 2013), which is also in line with the results of the present study, since conditions that achieved the highest embodiment were those with the highest visuotactile congruency based on prior knowledge, i.e., seeing an avatar or feather touching the virtual body, while simultaneously perceiving a real human or a feather stroke. A critical aspect in relation to these past studies is that the authors only administered slow stroking affective touch on hairy skin, since until recently it was thought that only non-glabrous skin was innervated with slow-conducting unmyelinated tactile fibers (i.e., CT afferents) which have shown to be critical in social bonding and for experiencing touch as pleasant (McGlone et al., 2012; Olausson et al., 2002). Conversely, this assumption has been challenged by recent evidence showing that highly pleasant experiences in response to touch might be also accomplished when being touched on glabrous skin (Ackerley et al., 2014; Watkins et al., 2021) and also by a recent microneurography study showing that CT afferents also innervate glabrous skin (Watkins et al., 2021). Therefore, the pleasure felt when touched on glabrous skin does not seem to be a simple by-product of top-down processing, but a much more complex process also implicating bottom-up information. In the present study, somatosensory stimulation was administered to glabrous skin (i.e., palm on the hand), and in accordance with this recent evidence we found that this tactile stimulation led to increased arousal and emotional response, which also seems to have strengthened embodiment in an artificial body. These results point out the need for further studies researching how embodiment varies depending on the parts of a virtual body where touch is delivered and the affective qualities of touch. The decision of applying touch on the palm of the hand in this study was based on the use of ultrasonic mid-air haptic stimulation, which has been shown to be best perceived and to induce a stronger body ownership illusion when administered in the palmar surface of the hand compared to the dorsal surface (Freeman et al., 2014; Paneva et al., 2020; Pittera et al., 2019; Salagean et al., 2021).

In the present study, we also found that participants' gender might play a role in affective touch perception in VR. Specifically, we found that female participants reported a stronger illusion of being caressed by a female avatar or feather when compared to male participants. In line with this result, Fusaro et al. (2019) also observed a stronger illusion of vicarious touch in female participants compared to male participants when being caressed by a female avatar instead of a male avatar. Further research should investigate the underlying mechanisms explaining these observed gender differences in body-related illusions, despite that it is quite likely that these differences are driven by top-down factors (i.e., cultural and social norms), rather than by bottom-up factors, since for instance there is no evidence that gender modulates phenomena such as mirror-touch synesthesia (Ward et al., 2018) or the strength of bodily illusions (Tacikowski et al., 2020).

Finally, we have also found that ultrasonic mid-air haptics might be a good candidate for simulating realistic affective touch in VR or social touch interactions in mediated communications. Although real skin-to-skin contact or physical contact with a feather led to the strongest level of illusory touch, the subjective ratings reported in relation to having the illusion of being touched or caressed in the conditions including mid-air ultrasonic tactile stimulation were still very high (median 6 in a 7-point Likert scale). These results have implications for studies and applications which aim to provide richer sensory feedback in immersive virtual environments designed for rehabilitation, education, or entertainment. This is a particularly relevant topic in light of the huge recent interest raised by the COVID-19 pandemic of adding touch interactions to VR experiences in order to improve technology mediated social experiences (Ahmed et al., 2016; Andersen & Guerrero, 2008; van Erp & Toet, 2015). Past studies have simulated social touch experiences in VR to assess whether it could increase compliance for other avatars' decisions (Bourdin et al., 2013; Harjunen et al., 2017; Świdrak & Pochwatko, 2019), increase human likeness of agents (Hoppe et al., 2020), or evoke different emotional reactions depending on the type of virtual social touch in multiuser interactions in VR (Sykownik & Masuch, 2020). However, none of these studies specifically focused on assessing the impact of virtual caresses based on providing different types of multisensory stimulation and being touched by different virtual representations. Based on the results of the present study, we propose that the addition of visuotactile feedback in VR scenarios simulating social interactions has enormous potential in areas such as psychological therapy (Kim et al., 2019), motor rehabilitation (Zakharov et al., 2020), pain treatment (Donegan et al., 2020; Matamala-Gomez et al., 2019), and VR-based videoconferences (Fermoselle et al., 2020), areas that should be researched in more detail in the near future.

The present study did not directly address some aspects that should be further researched by future studies. First, we only investigated the impact of being touched by a female avatar compared to a virtual inanimate object (i.e., feather), however we did not include a condition where participants were touched by a male avatar. Although this has already been researched by Fusaro et al. (2021) in a study where participants perceived vicarious touch in VR either delivered by a female or male avatar, that study did not include additional tactile stimulation. Interestingly, Fusaro et al. (2021) found that vicarious touch of a female avatar in intimate areas results in higher arousal than being touched by a male avatar, further corroborating that the gender of the toucher is a top-down factor that modulates reactions to virtual touches. Moreover, we only used mid-air ultrasonic tactile stimulation to artificially simulated touch, despite the existence of other haptic technologies (i.e., pin arrays, mechanical vibrators, water flow, among others). Future research should investigate whether such haptic technologies can also be used to effectively simulate affective touch and how they compare to ultrasonic airwaves. Moreover, to match as closely as possible the visual animations of being touched in VR with the ultrasonic mid-air tactile stimulation we had to calibrate the ultrasound-based haptics for each hand size, resulting in slight variations of the velocities of touch experienced by each participant. However, velocities were always in the range of what is considered slow velocity affective touch by past research (i.e., 1 to 10 cm/s). It would be interesting that future studies investigate if the present results vary when using constant velocities, as well as if there are critical velocities ranges for touch to be considered pleasant or unpleasant in VR. Finally, as already noted, in the present study we only stimulated the glabrous skin of the palmar side of the hand due to its higher tactile sensitivity, however further research is needed to better understand whether bodily-related illusions (i.e., illusion of being touched and embodiment) are differently impacted when participants are touched on different body parts with glabrous and non-glabrous skin. Moreover, in this study we researched the impact of being caressed in VR without the inclusion of a control condition based on the perception of nociceptive stimuli or neutral non-affective touch. It would be beneficial for future studies to include such types of control conditions and compare them to being caressed in VR while simultaneously perceiving congruent visuotactile feedback.

## 5. Conclusion

The present study aimed to better understand the emotional, behavioral and physiological correlates of being virtually caressed in VR either by an avatar or inanimate virtual object (feather), depending on the type of sensory feedback provided. Our results indicate that congruent spatiotemporal visuotactile feedback, compared to single visual feedback, boosts the illusion of being affectively touched in VR, as well as increases embodiment, pleasantness, arousal, and physiological response. This was true for the administration of tactile stimulation, based on ultrasound waves or real physical interpersonal contact, while simultaneously observing the touch of an avatar or virtual feather on an embodied virtual hand. However, the highest levels of illusory affective touch, pleasantness, and arousal, were evoked by conditions including real interpersonal contact (i.e., skin-to-skin or physical feather) compared to the other experimental conditions. With respect to being touched by an avatar or feather, we show that arousal and the desire to withdraw the caressed virtual hand is highest when touched by an avatar compared to an inanimate object. Finally, female participants reported a stronger illusion of being caressed in VR when compared to male participants. This study further contributes to the understanding of the emotional and physiological impact of affective touch in VR depending on factors such as sensory feedback provided and the virtual representations that touch a user in an immersive virtual environment. Moreover, it also highlights new opportunities for the design of richer and more realistic social interactions in VR, finding that mid-air ultrasonic stimulation might be a good candidate to simulate relatively realistic affective touch in VR, despite not reaching the same level of realism when compared to real interpersonal touch.

**Declaration of interest:** None

**Funding:** This research has received funding from the European Union's Horizon 2020 research and innovation programme under grant agreement #737087 (Levitate).

**Authors Contributions:** Conceived, designed, and supervised the experiments: S.S. and J.J. Implemented Technical Setup and Run Experiments: I.S. Analyzed the data: S.S. Wrote the paper: S.S., with the help of J.J and I.S.

Etzi, R., Zampini, M., Juravle, G., & Gallace, A. (2018). Emotional visual stimuli affect the evaluation of tactile stimuli presented on the arms but not the related electrodermal responses. *Experimental Brain Research 2018 236:12*, *236*(12), 3391–3403. https://doi.org/10.1007/S00221-018-5386-0

Fermoselle, L., Gunkel, S. N., ter Haar, F., Dijkstra-Soudarissanane, S., Toet, A., Niamut, O., & van der Stap, N. (2020). Let's Get in Touch! Adding Haptics to Social VR. *ACM International Conference on Interactive Media Experiences*. https://doi.org/10.1145/3391614

Field, T. (2010). Touch for socioemotional and physical well-being: A review. In *Developmental Review* (Vol. 30, Issue 4, pp. 367–383). Academic Press. https://doi.org/10.1016/j.dr.2011.01.001

Freeman, E., Brewster, S., & Lantz, V. (2014). Tactile feedback for above-device gesture interfaces: Adding touch to touchless interactions. *ICMI 2014 - Proceedings of the 2014 International Conference on Multimodal Interaction*, 419–426. https://doi.org/10.1145/2663204.2663280

Fusaro, M., Tieri, G., & Aglioti, S. M. (2016). Seeing pain and pleasure on self and others: behavioral and psychophysiological reactivity in immersive virtual reality. *Https://Doi.Org/10.1152/Jn.00489.2016*, *116*(6), 2656–2662. https://doi.org/10.1152/JN.00489.2016

Fusaro, M., Tieri, G., & Aglioti, S. M. (2019). Influence of cognitive stance and physical perspective on subjective and autonomic reactivity to observed pain and pleasure: An immersive virtual reality study. *Consciousness and Cognition*, *67*, 86–97. https://doi.org/10.1016/j.concog.2018.11.010

Fusaro, Martina, Lisi, M. P., Tieri, G., & Aglioti, S. M. (2021). Heterosexual, gay, and lesbian people's reactivity to virtual caresses on their embodied avatars' taboo zones. *Scientific Reports*, *11*(1), 2221. https://doi.org/10.1038/s41598-021-81168-w

Fusaro, Martina, Lisi, M., Tieri, G., & Aglioti, S. M. (n.d.). *Touched by vision: how heterosexual, gay, and lesbian people react to the view of their avatar being caressed on taboo body parts*. https://doi.org/10.31234/OSF.IO/DKZJ5

Gonzalez-Franco, M., & Peck, T. C. (2018). Avatar Embodiment. Towards a Standardized Questionnaire. *Frontiers in Robotics and AI*, *5*, 74. https://doi.org/10.3389/frobt.2018.00074

Haans, A., & IJsselsteijn, W. (2006). Mediated social touch: A review of current research and future directions. In *Virtual Reality* (Vol. 9, Issues 2–3, pp. 149–159). Springer. https://doi.org/10.1007/s10055-005-0014-2

Harjunen, V. J., Spapé, M., Ahmed, I., Jacucci, G., & Ravaja, N. (2017). Individual differences in affective touch: Behavioral inhibition and gender define how an interpersonal touch is perceived. *Personality and Individual Differences*, *107*, 88–95. https://doi.org/10.1016/j.paid.2016.11.047

Hazem, N., Beaurenaut, M., George, N., & Conty, L. (2018). Social Contact Enhances Bodily Self-Awareness. *Scientific Reports*, *8*(1), 1–10. https://doi.org/10.1038/s41598-018-22497-1

Hoppe, M., Rossmy, B., Neumann, D. P., Streuber, S., Schmidt, A., & MacHulla, T. K. (2020). A Human Touch: Social Touch Increases the Perceived Human-likeness of Agents in Virtual Reality. *Conference on Human Factors in Computing Systems - Proceedings*, *20*. https://doi.org/10.1145/3313831.3376719

Huisman, G. (2017). Social Touch Technology: A Survey of Haptic Technology for Social Touch. *IEEE Transactions on Haptics*, *10*(3), 391–408. https://doi.org/10.1109/TOH.2017.2650221

Jakubiak, B. K., & Feeney, B. C. (2017). Affectionate Touch to Promote Relational, Psychological, and Physical Well-Being in Adulthood: A Theoretical Model and Review of the Research. *Personality
19

Serino, A., Alsmith, A., Costantini, M., Mandrigin, A., Tajadura-Jimenez, A., & Lopez, C. (2013). Bodily ownership and self-location: Components of bodily self-consciousness. *Consciousness and Cognition*, *22*(4), 1239–1252. https://doi.org/10.1016/J.CONCOG.2013.08.013

Shin, M., Kim, S. J., & Biocca, F. (2019). The uncanny valley: No need for any further judgments when an avatar looks eerie. *Computers in Human Behavior*, *94*, 100–109. https://doi.org/10.1016/j.chb.2019.01.016

Świdrak, J., & Pochwatko, G. (2019). Being touched by a virtual human. *Proceedings of the 19th ACM International Conference on Intelligent Virtual Agents*. https://doi.org/10.1145/3308532

Sykownik, P., & Masuch, M. (2020). The Experience of Social Touch in Multi-User Virtual Reality. *26th ACM Symposium on Virtual Reality Software and Technology*. https://doi.org/10.1145/3385956

Tacikowski, P., Fust, J., & Ehrsson, H. H. (2020). Fluidity of gender identity induced by illusory body-sex change. *Scientific Reports*, *10*(1), 14385. https://doi.org/10.1038/s41598-020-71467-z

Tieri, G., Tidoni, E., Pavone, E. F., & Aglioti, S. M. (2015). Body visual discontinuity affects feeling of ownership and skin conductance responses. *Scientific Reports*, *5*(1), 17139. https://doi.org/10.1038/srep17139

van Dooren, M., de Vries, J. J. G. G. J., & Janssen, J. H. (2012). Emotional sweating across the body: Comparing 16 different skin conductance measurement locations. *Physiology and Behavior*, *106*(2), 298–304. https://doi.org/10.1016/j.physbeh.2012.01.020

van Erp, J. B. F., & Toet, A. (2015). Social Touch in Human-Computer Interaction. *Frontiers in Digital Humanities*, *2*, 2. https://doi.org/10.3389/fdigh.2015.00002

Ward, J., Schnakenberg, P., & Banissy, M. J. (2018). The relationship between mirror-touch synaesthesia and empathy: New evidence and a new screening tool. *Cognitive Neuropsychology*, *35*(5–6), 314–332. https://doi.org/10.1080/02643294.2018.1457017

Watkins, R. H., Dione, M., Ackerley, R., Wasling, H. B., Wessberg, J., & Löken, L. S. (2021). Evidence for sparse C-tactile afferent innervation of glabrous human hand skin. *Https://Doi.Org/10.1152/Jn.00587.2020*, *125*(1), 232–237. https://doi.org/10.1152/JN.00587.2020

Wilson, G., Carter, T., Subramanian, S., & Brewster, S. (2014). Perception of ultrasonic haptic feedback on the hand: Localisation and apparent motion. *Conference on Human Factors in Computing Systems - Proceedings*, 1133–1142. https://doi.org/10.1145/2556288.2557033

Zakharov, A. V., Bulanov, V. A., Khivintseva, E. V., Kolsanov, A. V., Bushkova, Y. V., & Ivanova, G. E. (2020). Stroke Affected Lower Limbs Rehabilitation Combining Virtual Reality With Tactile Feedback. *Frontiers in Robotics and AI*, *0*, 81. https://doi.org/10.3389/FROBT.2020.00081
21